\documentclass[aps,pra,showpacs,floatfix,twocolumn,superscriptaddress]{revtex4-1}
\usepackage{graphicx}
\usepackage{bm,amsmath}
\usepackage{dcolumn}
\newcommand{\be}{\begin{equation}}
\newcommand{\ee}{\end{equation}}
\newcommand{\bea}{\begin{eqnarray}}
\newcommand{\eea}{\end{eqnarray}}
\newcommand{\la}{\langle}
\newcommand{\ra}{\rangle}
\begin{document}
\title{Extended Bose-Hubbard model with pair hopping  on the triangular 
 lattice}

\author{Yancheng Wang}
\affiliation{ Physics Department, Beijing Normal University, Beijing 100875, China}
\author{Wanzhou Zhang}
\affiliation{ College of Physics and Optoelectronics, Taiyuan University of Technology Shanxi 030024, China}
\author{Hui Shao}
\affiliation{ Physics Department, Beijing Normal University, Beijing 100875, China}
\author{Wenan Guo}
\email{Corresponding author: waguo@bnu.edu.cn}
\affiliation{ Physics Department, Beijing Normal University, Beijing 100875, China}

\date{\today}
\begin{abstract}
We study systematically an extended Bose-Hubbard model on the triangular
lattice by means of a meanfield method based on the Gutzwiller ansatz.
Pair hopping terms are explicitly included and a three-body constraint
is applied. The zero-temperature phase diagram and a variety of quantum phase
transitions are investigated in great detail.  In particular, we 
show the existence and stability of the pair supersolid phase. 

\end{abstract}
\maketitle

\section{Introduction}
\label{sec:intro}
The development in experimentally manipulating ultra-cold atoms in an
optical lattice has allowed the realization of novel quantum states
and quantum phase transition in strongly correlated systems \cite{qgas},
e.g., a quantum phase transition from a superfluid (SF) to a Mott insulator (MI)
has been predicted and observed \cite{Greiner}.
The condensation of paired electrons, which provides the basis of
superconductivity of metallic superconductor,
plays an essential role in modern condensed-matter physics.
Thus realizing pairing related novel quantum states in the context of
ultra-cold atoms has attracted considerable recent interest, both in
theoretical and experimental research.

It has been demonstrated recently that such states can be realized for 
lattice bosons with attractive on-site interactions, which is stabilized by a 
three-body constraint \cite{zoller1, th1}. The three-body constraint has been 
realized by large three-body loss processes\cite{3bodyloss, 3body2}. The 
system can be mapped into spin-1 atoms at unit filling\cite{sp1mott}.
Besides the conventional atomic superfluid (ASF) phase,
a pair (dimer) superfluid (PSF) phase consisting of the
condensation of boson pairs emerges under sufficiently strong
attraction\cite{zoller1,th1}.
The PSF state is manifested as a second-order processes of the single-atom 
hopping in
the optical lattice.  Various phase transitions among  the ASF, MI and PSF
are investigated in great detail \cite{zoller1, th1, mfy1,mfy2,psf,ycc,fintemp}.
Both the ground-state and the thermal phase diagrams are obtained.
However, the single-species pair supersolid (PSS) was not found in the present 
system when the nearest-neighbor repulsion is included,
except for an isolated continuous supersolid at the Dirac point \cite{zoller1}.
The three-body constraint and the effective pair hopping makes the system 
resemble hardcore bosons when onsite repulsion $U$ is weak. 
Thus the PSS state might suffer from the same instability of the
supersolid state (SS) for hardcore bosons on the square lattice
\cite{ssm3, sshcsq}.
The pair supersolid state was predicted only
in system with correlated hopping \cite{Schmidt, hcj} and in bilayer system
\cite{Trefzger}.

For hardcore bosons on the triangular lattice, 
supersolid state emerges basing on an order-by-disorder mechanism, by
which a quantum system avoids classical frustration\cite{tri1,tri2,tri3,tri4}.
Aiming to realize PSS state, we thus focus on lattice bosons on the 
triangular optical lattice. 
The system we consider is an extended Bose-Hubbard model  
with the three-body constraint $a_i^{\dag3}\equiv0$.
The pair hopping terms are explicitly included. 
The Hamiltonian is
\bea
H &=&-t \sum\limits_{\la i,j \ra} (a_{i}^{\dag }a_{j}+a_{j}^{\dag}a_{i}) 
-t_p\sum\limits_{\la i,j \ra} (a_{i}^{\dag }a_{i}^{\dag}a_{j}a_{j}+a_{j}^{\dag}a_{j}^{\dag }a_{i}a_{i}) 
\nonumber \\
&&  +\frac{U}{2}\sum\limits_{i}n_{i}(n_{i}-1)+V\sum\limits_{\la i,j \ra} n_{i}n_{j}-\mu\sum\limits_{i}n_{i} ,
\label{H}
\eea
where $\la i,j \ra$  denotes nearest-neighbor sites,
$a_i^{\dag} (a_i)$ is the boson creation (annihilation) operator at site $i$, 
and $n_i=a_i^{\dag}a_i$  the boson number operator;
$t$~$(t_p)$ is the single-atom (pair) hopping amplitude,
$U$  the on-site interaction,  $\mu$ the chemical potential, and
$V$ the nearest-neighbor repulsion. 
Such a system can be realized
experimentally for dipolar bosons polarized by an external electric field
and confined in an optical lattice \cite{Sowinski}. 
The pure pair hopping limit ($t=0$) can be realized by a 
mechanism based on collisions that induce transport\cite{cor1}. 
At the limit $V=0$, the system can be realized in an atom-molecule
coupling system in a state-dependent optical lattice \cite{xfzhou}.
A similar effective model of bosons with the three-body constraint 
can be realized with spin-1 atoms\cite{sp1mott}.

In present work we systematically study the model by means of 
the meanfield method based on the Gutzwiller ansatz\cite{Sheshadri,Jaksch,Iskin, Lu}.  
The zero-temperature phase diagram of the system is studied in great detail.
The existence of the PSS phase at the $t \to 0$ limit is explained by a 
mapping between the present model and the hardcore bosons on the triangular 
lattice. We then show that the PSS phase is stable in the presence of small 
on-site repulsion $U$ and single-atom hopping $t$.

This paper is organized as follows: we first discuss the classical limit
of model (\ref{H}) at zero temperature in Sec.\ref{sec:class}. 
The model shows various solid states, 
which are the basis to form a pair supersolid state. 
Then we describe the mean-field method  
in Sec. \ref{sec:meanfield}. 
We present our main results in Sec. \ref{sec:res}. 
We discuss the the noninteracting case ($U=V=0$) in Sec. \ref{sec:u0v0}.   
For interacting cases, 
we present the zero-temperature phase diagram 
in the limit single-atom hopping $t=0$,
focusing on the parameter region where the pair supersolid phase emerges, 
in Sec. \ref{sec:t=0}. 
We then investigate if the results are stable against putting a
finite $t/t_p$ in Sec. \ref{sec:tvstp}, in which
a more realistic phase diagram is presented.
We conclude in Sec. \ref{sec:conc}.

\section{Classical Limit}
\label{sec:class}
\begin{figure}[htbp]
\vskip 0.8cm
\includegraphics[width=4cm]{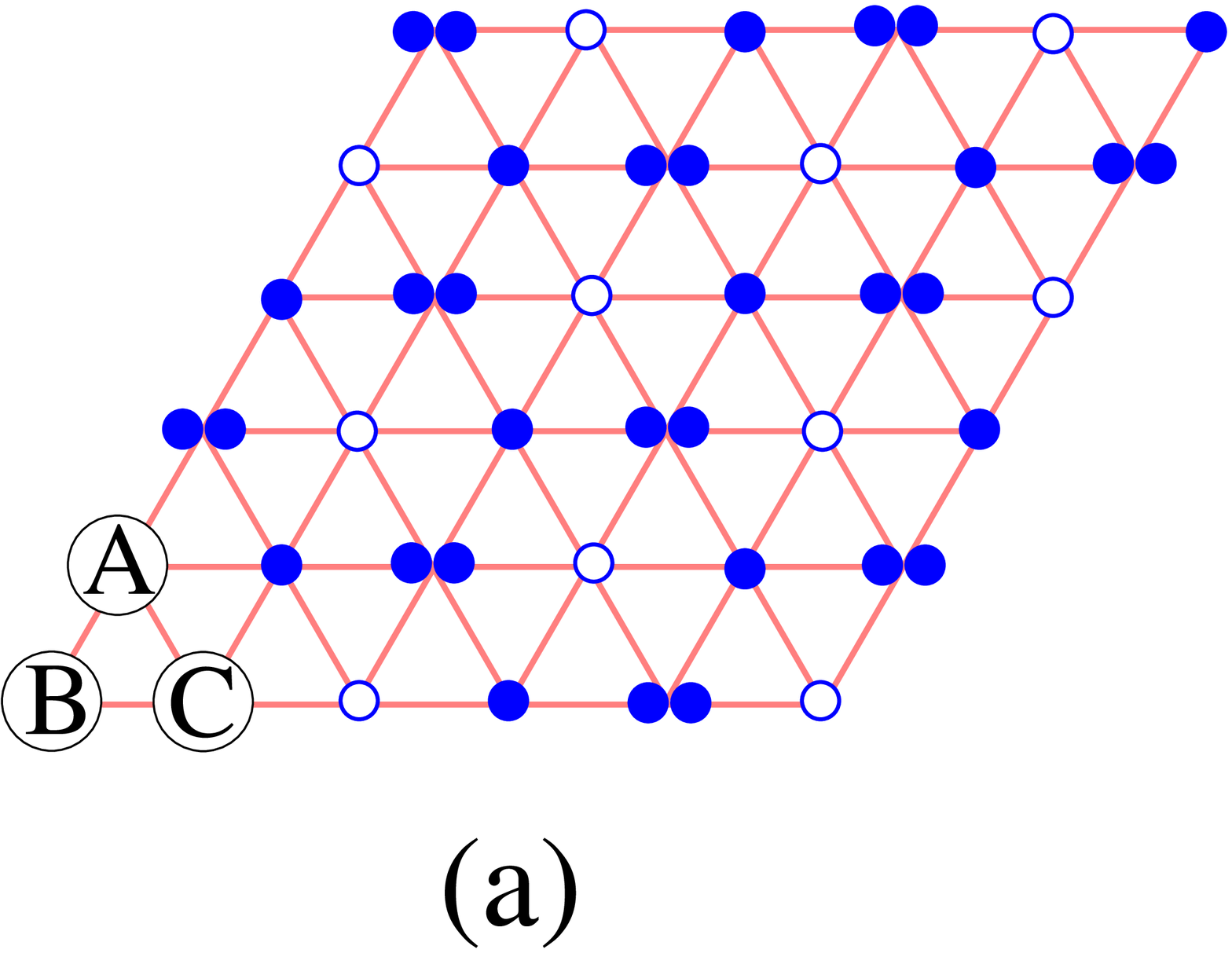}
\includegraphics[width=4cm]{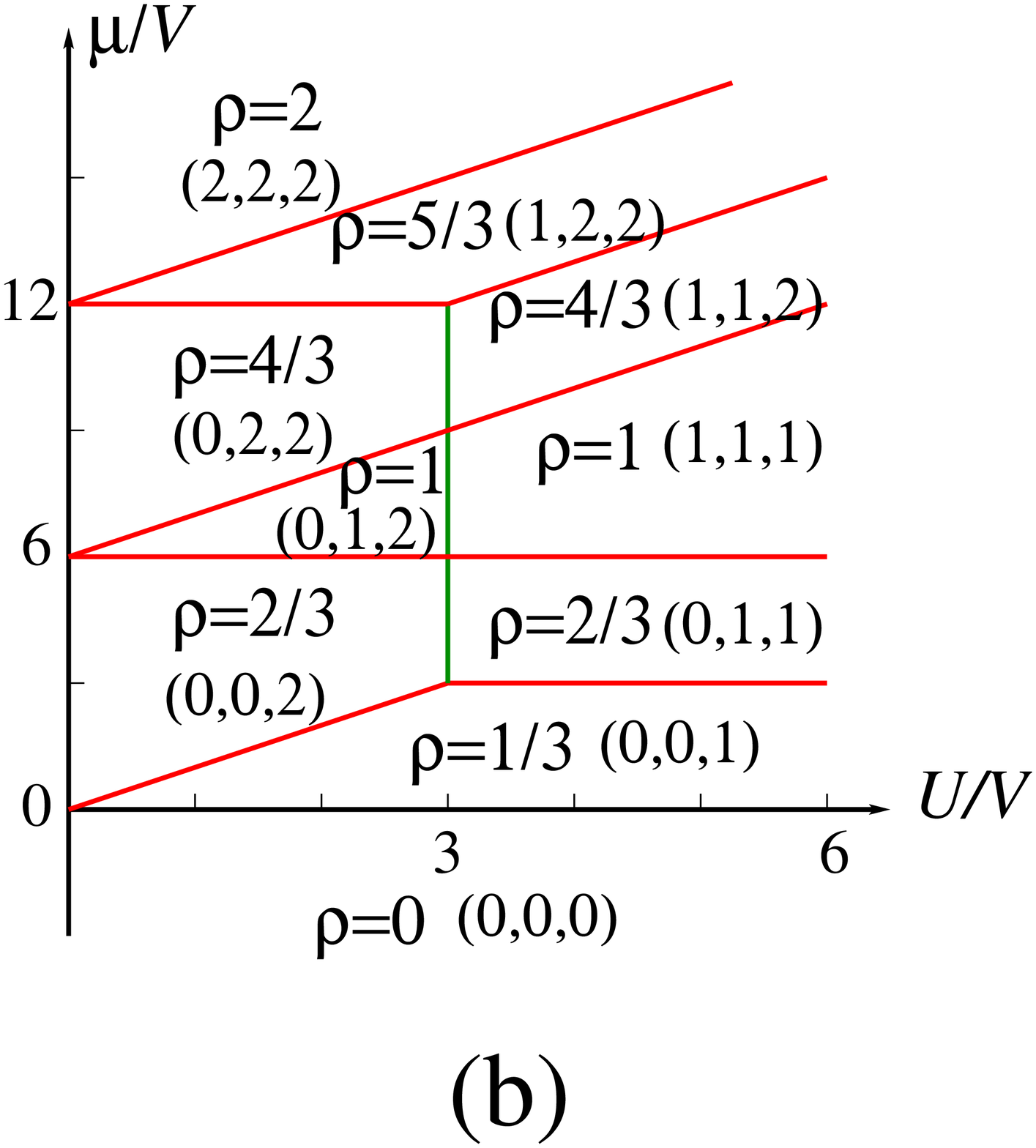}
\caption{(Color online) (a) The triangular lattice and its three sublattices.
The $(0, 1, 2)$ solid is also shown.
(b)Zero-temperature phase diagram  
in the classical limit  $t=0$, $t_p=0$.}
\label{t0tp0}
\end{figure}

There are three sublattices $A, B$ and $C$ in the triangular lattice,
as shown in Fig. \ref{t0tp0} (a).
In the classical limit ($t=0,t_p=0$) and zero temperature, the energy per unit 
cell is 
\bea
\label{eqs2}
&E_{\triangle}& = \frac{U}{2}( n_A  ( n_A -1)+ n_B ( n_B  -1) 
+n_C ( n_C -1)) \nonumber \\
&-&\mu( n_A + n_B  + n_C  )
+\frac{zV}{2} ( n_A   n_B  + n_B  n_C + n_C  n_A ),
\eea
where $z=6$ is the coordination number of the triangular lattice.
$n_A, n_B, n_C$  is the occupation number on the sublattice $i=A, B$,
and $C$, respectively.  

A solid state is formed when the symmetry of occupation on the three
sublattices is broken spontaneously. 
The order can be represented by the occupation pattern 
$(n_A, n_B, n_C)$, with $n_A, n_B, n_C$ not equal. If $n_A=n_B=n_C$,
the pattern represents an MI state.

By comparing the energy per unit cell, we obtain the phase boundaries between 
various ordered phases, as shown in Fig. \ref{t0tp0}(b).
The solid states $(0, 0, 2)$, $(0, 1, 2)$, $(0, 2, 2)$, $(1, 1, 2)$ and 
$(1, 2, 2)$, 
which have two bosons sitting on one or two sublattices,
are of special interest. 
The corresponding density is $\rho=2/3, 1, 4/3, 4/3, 5/3$, 
respectively. 
Other solid  states related with  patterns $(0, 0, 1), (0, 1, 1)$ and  
MI states, $(0, 0, 0), (1, 1, 1), (2, 2, 2)$, with corresponding densities 
$\rho=0, 1, 2$, are also found.

\section{Meanfield Method}
\label{sec:meanfield}
We start with the Gutzwiller ansatz 
of the many-body wave function
\be
|\psi \ra = \prod_i (\sum_n^{n_{max}} c_{i, n}|i, n \ra),
\label{GA}
\ee
where $|i,n\ra$ is the Fock state of $n$ bosons occupying the site $i$, and
$c_{i,n}$ is the probability amplitude. 
The three-body constraint 
requires the maximum number of bosons  on each site $n_{max}=2$. 
The mean-field decoupling approximation
\bea
a_{i}^{\dag}a_{j}&=&\la a_{i}^{\dag}\ra a_{j}+a_{i}^{\dag}\la a_{j}\ra - \la a_{i}^{\dag}\ra \la a_{j}\ra \, , \\
a_{i}^{\dag}a_{i}^{\dag}a_{j}a_{j}&=&\la a_{i}^{\dag2}\ra a_{j}^2+a_{i}^{\dag2}\la a_{j}^{2}\ra -\la a_{i}^{\dag2}\ra \la a_{j}^{2}\ra \, , \\
n_{i}n_{j}&=&\la n_{i}\ra n_{j}+n_{i}\la n_{j}\ra -\la n_{i}\ra \la n_{j} \ra
\eea
are exact for such a state.
The atomic superfluid order parameter is defined as 
$\phi_{a,i} \equiv \la a_{i}\ra 
= c^*_{i,0} c_{i,1}+\sqrt{2}c^*_{i,1}c_{i,2}$ and the 
pair superfluid order parameter is $\phi_{p,i} \equiv \la a_{i}^{2}\ra=\sqrt{2} c^*_{i,0}c_{i,2} $.
The average occupancy  $\la n_i \ra=\sum_n^{2}n |c_{i,n}|^2$.
For simplicity, we choose the order parameters to be real, i.e., 
$\phi_{a(p),i}=\phi_{a(p),i}^{*}$.
The meanfield version of the Hamiltonian (\ref{H}) is thus written  as a 
sum over single-site terms
$H=\sum_i H_i$, with
\bea
H_i&=&-t (\bar{\phi}_{a,i}a_{i}^{\dag}+ {\rm H.c.})  
      -t_p (\bar{\phi}_{p,i}a_{i}^{\dag2} + {\rm H.c.}) \nonumber\\
&&    +\frac{U}{2}n_{i}(n_{i}-1)+V (n_i \bar{n}_i- {\la n_i \ra \bar{n}_i\over 2} ) -\mu n_i\\
&&+t \bar{\phi}_{a,i} {\phi}_{a,i} +t_p \bar{\phi}_{p,i}{\phi}_{p,i}, \nonumber
\eea
where $\bar{\phi}_{a(p),i}=\sum_{\la j\ra_i}\phi_{a(p), j}$ and 
$\bar{n}_i=\sum_{\la j\ra_i} \la n_{j} \ra$ 
sum over sites $j$ neighboring to site $i$.
Written in the matrix form, 
\bea
H_i=\left(
\begin{array}{ccc}
d(0)                        &-t \bar{\phi}_{a,i} & -\sqrt{2} t_p \bar{\phi}_{p,i} \\
-t \bar{\phi}_{a,i}           &d(1)              & -\sqrt{2} t \bar{\phi}_{a,i} \\  
-\sqrt{2} t_p \bar{\phi}_{p,i} &-\sqrt{2} t \bar{\phi}_{a,i}   &d(2)
\end{array} 
\right),
\label{matrix}
\eea
where the diagonal element 
\bea
d(k) &=& \frac{U}{2}k(k-1) + V (k-\frac{\la n_i \ra}{2}) \bar{n}_i -\mu k 
\nonumber \\
     &+& t \bar{\phi}_{a, i}\phi_{a, i} + t_p \bar{\phi}_{p, i}\phi_{p, i} \, ,
\label{dk}
\eea
with $k=0,1,2$.

Without the nearest-neighbor repulsion ($V=0$), there is no solid ordering. 
The Gutzwiller ansatz (\ref{GA}) reduces to
\be
|\psi \ra = (c_0 |0 \ra + c_1 |1\ra + c_2 |2 \ra )^N,
\ee
with $N$ the number of lattice sites. Therefore,  
$\phi_{a(p),i}=\phi_{a(p)}, \la n_i \ra =\la n \ra$,
and $\bar{\phi}_{a(p),i}=z \phi_{a(p)}, \bar{n}_i=z \la n \ra$.
We then find the ground state in a self-consistent way:
Given an initial state $|\psi \ra $, the order parameters $\phi_{a(p)}$ and 
the occupancy $\la n \ra$ are calculated. 
The matrix (\ref{matrix}) is thus obtained and diagonalized. 
$\phi_{a(p)}$ and $\la n \ra$ are then evaluated again from 
the ground state and put back to (\ref{matrix}). 
This is done iteratively until the 
order parameters converge to a self-consistent solution.
There can be more than one self-consistent solutions. 
The one with lowest energy is chosen. This procedure is equivalent to
minimization of the ground energy with respect to the order parameters
\cite{Sheshadri}.

With the nearest-neighbor repulsion ($V \neq 0$) turning on, 
a solid order may emerge.  To accommodate the solid order, we assume  
\be
|\psi \ra 
              =\prod_{i=A,B,C}(c_{i,0} |i, 0 \ra + c_{i, 1} |i, 1\ra + c_{i, 2} |i, 2\ra )^{N_i},
\ee
where $N_i$ is the number of sites in sublattice $i \in A, B, C$. 
The solid order is represented by the occupation pattern 
($\la n_A \ra, \la n_B \ra, \la n_C \ra$), which can be depicted by a 
solid order parameter  
$\Delta \rho^2 \equiv \sum\limits_{i}( \la n_i \ra -\rho)^2$,
with $\rho=\frac{1}{3} \sum\limits_{i} \la n_i \ra $
the density of bosons.
For a site in sublattice $A$, one has  
\bea
\bar{\phi}_{a(p),A}&=&\frac{z}{2}(\phi_{a(p),B}+\phi_{a(p),C}), \nonumber \\
\bar{n}_{A}&=&\frac{z}{2}(\la n_{B}\ra +\la n_{C}\ra ). 
\eea
Their equivalences under cyclic transformation can also be found. 
The ground state can be found self-consistently.
We start with an arbitrary state, from which
the parameters $\phi_{a(p),A}$, $\phi_{a(p),B}$, $\phi_{a(p),C}$, 
$\la n_A \ra $, $\la n_B \ra $, and $\la n_C \ra $ are 
evaluated. The matrix (\ref{matrix}) for sublattice $A$ is first 
obtained and diagonalized. 
From the obtained ground state, ``new" $\phi_{a(p), A}$ and $\la n_A \ra$ are
calculated.  Then the matrix (\ref{matrix}) for sublattice $B$ is obtained
and diagonalized. ``new" $\phi_{a(p), B}$ and $\la n_B \ra$ are obtained and
used in the matrix (\ref{matrix}) for sublattice $C$. 
This is done recursively, until the estimated order parameters and 
occupancies converge. Again, the lowest energy solution is chosen. 

The zero-temperature phase diagram of the system is constructed by checking 
the obtained solutions. 
The MI and solid states are characterized by $\phi_{a}=\phi_p=0$, and 
$\Delta \rho^2 =0$ and $\Delta \rho^2 \ne 0$, respectively.
The ASF and atomic super-solid (ASS) phases are characterized by 
$\phi_{a} \ne 0, \phi_p \ne 0$, and $\Delta \rho^2=0$ and 
$\Delta \rho^2 \ne 0$, respectively. 
The PSF and PSS both have the pair superfluid character $\phi_{a}=0$ and 
$\phi_{p} \ne 0$.  $\Delta \rho^2=0$ in the PSF state,  
but $\Delta \rho^2 \ne 0$ in the PSS phase. In the PSS state,
$\phi_{p,A}, \phi_{p,B}$ and $\phi_{p,C}$ are different in general. We take 
the mean value $\phi_p \equiv \sum_i \phi_{p,i}/3$ as the order parameter.

\section{Results}
\label{sec:res}
In this section, we present our main results.
\subsection{Competition between $t$ and $t_p$ in the non-interacting case }
\label{sec:u0v0}
\begin{figure}[htbp]
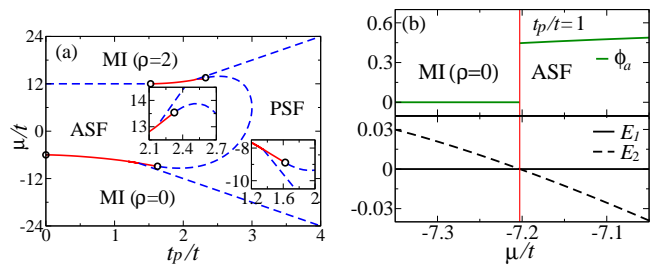

\begin{center}
\includegraphics[width=4.2cm ]{tp-mu-u0-v0.eps}~~
\includegraphics[width=4.0cm ]{t1-tp1-U0-V0.eps}
\end{center}
\caption{(Color online)
(a) Zero-temperature phase diagram at $U=0$, $V=0$.
The dashed (blue) lines and the black dot represent continuous phase 
transitions. The solid
(red) lines represent first order transitions. The open circles denote the 
tricritical points. The insets show the details around the 
ASF-PSF tricritical points.
(b) Levelcrossing and the ASF order parameter $\phi_a$ jump at the MI 
($\rho=0)$-ASF transition point, along the $t_p/t=1$ line.  
}
\label{ph0}
\end{figure}

We start with the non-interacting limit,  focusing on the competition between
the single-atom hopping $t$ and the pair hopping $t_p$. 
Without $U$ and $V$, the diagonal element (\ref{dk}) of (\ref{matrix}) further 
reduces to
\be
d(k)=-\mu k+ z t \phi_a^2+ z t_p \phi_p^2.
\ee
The meanfield solutions for various $\mu/t$ and $t_p/t$ are found in the way 
described in Sec. \ref{sec:meanfield} for $V=0$ case.
The phase diagram is constructed by checking the obtained solutions,
as shown in Fig. \ref{ph0} (a).
There is a $\rho=0$ MI (empty) phase and a $\rho=2$ MI phase at negative and
large chemical potentials, respectively. The MI phase is the result of the 
three-body constraint.
At small pair hopping, there is an ASF phase between the empty phase and the MI 
phase.  The transitions between the ASF phase and the two MI phases can be
continuous or first order, depending on the ratio $t_p/t$. 
This is because, in presence of $t_p$, an energylevel crossing may preempt the
continuous evolving of the ground state. 
The first order transition behavior is shown in Fig. \ref{ph0} (b),
in which two energylevels cross each other,
leading to a discontinuous change of order parameters.
Similar MI-ASF transition behaviors are found in the attractive Bose-Hubbard 
model with three-body constraint \cite{mfy1}. 
As pair hopping $t_p$ becomes large enough, an ASF-PSF transition
happens as expected. 
The transition is found to be first order, or continuous, depending $t_p/t$, 
as illustrated in Fig. \ref{ph0} (a).
Several tricritical points where the jump of order parameters vanishes are 
found. 
However, according to \cite{zoller1,mfy2}, the coupling of the Ising order 
parameter and the Goldstone mode will drive this quantum phase transition to 
be a weakly first-order one through the Coleman-Weinberg mechanism \cite{CW}. 
For such a weak first-order transition the mean-field theory may be not 
reliable.

\vskip 0.1cm

\subsection{ The limit single-atom hopping $t=0$}
\label{sec:t=0}

We now turn to the interacting case. To demonstrate the effect
of the pair hopping term, we first discuss
the limit $t=0$,  at which Eq. (\ref{matrix}) reduces to
\bea
H_i=\left(
\begin{array}{ccc}
d(0)                        &0                   & -\sqrt{2} t_p \bar{\phi}_{p,i} \\
0                           &d(1)                & 0  \\
-\sqrt{2} t_p \bar{\phi}_{p,i} &0                &d(2)
\end{array}
\right),
\label{t0matrix}
\eea
where the diagonal elements read
\be
d(k)=\frac{U}{2}k(k-1) + V (k-\frac{\la n_i \ra}{2}) \bar{n}_i - \mu k+ t_p \bar{\phi}_{p, i}\phi_{p, i} ,
\ee
with $k=0,1,2$.
Without $t$, the Hamiltonian bears the particle-hole symmetry, 
which means that it is invariant under the transformation $n \to 2-n$ 
and $\mu \to 2 z V -\mu +U$. 

\subsubsection{ $t=0$, $U \neq 0$, $V=0$}
With the on-site repulsion $U$ present, 
the PSF state is the only possible superfluid state and a $\rho=1$ MI state 
emerges.  
Without the nearest-neighbor interaction $V$, no solid order presents. 
$\phi_{p,i}=\phi_p$ is uniform and $\bar{\phi}_{p_i}=z \phi_p$.
Competition between the on-site repulsion and the pair hopping leads to phase 
transitions among MI and PSF phases.

The diagonal terms in Eq. (\ref{t0matrix}) further reduces to 
$d(k)=\frac{U}{2}k(k-1) - \mu k+ z t_p \phi_p^2$, and the 
three eigenvalues are found as follows
\bea
E_{1,2} &=& z t_p \phi_p^2+\frac{U}{2}-\mu \mp \sqrt{(\frac{U}{2}-\mu)^2 +2(z t_p)^2 \phi_p^2}, \nonumber\\ 
E_3& =& z t_p \phi_p^2-\mu. 
\eea

$E_1$ is the ground state if $\mu<0$ or $\mu>U$, considering $U>0$.
Note that $\phi_p$ is self-consistently determined by the 
eigenvector $\vec{v}_1=(c_0, 0, c_2)$ associated to $E_1$
as $\phi_p=\sqrt{2} c_2 c_0$ and the eigenvector associated to $E_3$ is 
$\vec{v}_3=(0, 1, 0) $.
Given parameters $t_p, \mu$ and $U$, the self-consistent solution $E_1(\phi_p)$ 
should satisfy the condition $\partial E_1/ \partial \phi_p=0$, which 
determines the superfluid parameter $\phi_p$ of the ground state:
\be
\phi_p=\left\{
   \begin{array}{ll}
   0                               &~~~((\frac{U}{2}-\mu)/zt_p)^2 \ge 1;\\
   \frac{1}{2}\sqrt{1-((\frac{U}{2}-\mu)/zt_p)^2}   &~~~((\frac{U}{2}-\mu)/zt_p)^2<1.
   \end{array}
   \right.
\label{phip}
\ee
Thus the PSF and the MI ($\rho=2$) boundary is  
$\mu/U-1/2=z t_p/U$, where $\vec{v}_1$ evolves to (0, 0, 1) and $\phi_p=0$.
The PSF and  the MI ($\rho=0$)  boundary is 
$\mu/U-1/2= -z t_p/U$, where $\vec{v}_1$ evolves to (1, 0, 0) and 
$\phi_p=0$.

If $0 \le \mu/U \le 1$, then the eigenlevels $E_1$ and $E_3$ may cross each 
other at a first order transition point. 
If both sides of the point has $\phi_p=0$, then $E_1=E_3$ yields 
$\mu/U=1$, which is the MI ($\rho=1$) and MI ($\rho=2$) transition line, or, 
$\mu/U=0$, which is the MI ($\rho=1$)  and MI ($\rho=0$) transition line. 
It is still possible that $\phi_p \ne 0$ on one side of the transition and 
$\phi_p=0$ on the other side, which leads to the following equation 
\be
z t_p \phi_p^2+\frac{U}{2}-\mu - \sqrt{(\frac{U}{2}-\mu)^2 +2(z t_p)^2 \phi_p^2}
=-\mu,
\ee
with $\phi_p \ne 0 $ given by Eq. (\ref{phip}). This gives the PSF and 
MI ($\rho=1$) transition line 
\be
2 zt_p/U=1+\sqrt{1-(2\mu/U-1)^2}.
\ee

\begin{figure}[htbp]
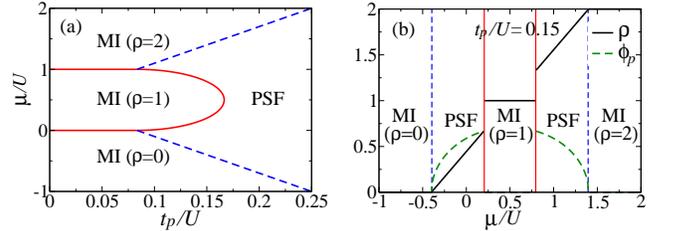

\vskip 0.1cm
\includegraphics[width=4.2cm]{tp-mu-u1-v0.eps}~~
\includegraphics[width=3.9cm]{tp0.15-u1-v0-cp.eps}
\caption{(Color online)
(a) Zero-temperature phase diagram at $t=0$, $V=0$.
The solid (red) lines denote first-order phase transitions,
while the dashed (blue) lines represent continuous phase transitions.
(b) Density $\rho$ and pair superfluid parameter $\phi_p$
versus $\mu/U$ at $t_p/U=0.15$.  }
\label{une0v0ph}
\end{figure}

The phase diagram is shown in Fig.~\ref{une0v0ph}(a), which can be compared 
with the similar phase diagram presented in Ref. \cite{xfzhou}, where the  
three-body constraint was not applied.  
Due to the particle-hole symmetry, the phase diagram is symmetric about 
$\mu/U=1/2$. 
The nature of phase transitions involved is further investigated, as 
illustrated in Fig.~\ref{une0v0ph}(b), where $\mu/U$ varies from -1 to 2 
along the line $t_p/U=0.15$. 
It is clear that the transitions between the $\rho=0, \rho=2$ MI phases 
and the PSF phase are continuous,
while the transition between the $\rho=1$ MI and the PSF is first order.
The latter behavior is different from the nature  of the SF-MI transition in 
the softcore Bose-Hubbard model with single-atom hopping. 
The $\rho=1$ MI state $\vec{v}_3=(0, 1, 0)$
can not evolve to the PSF state $\vec{v}_1$ continuously in the truncated 
Hilbert space due to the three-body constraint. 

\subsubsection{$t=0$, $U=0$, $V \ne 0$}
\label{sec:u0vne0}
\begin{figure}[tpbh]
\vskip 0.4cm
\includegraphics[width=7cm]{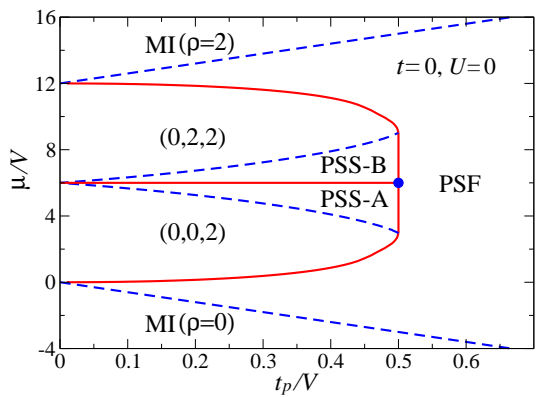}
\caption{(Color online)
Zero-temperature phase diagram at $t=0, U=0$.
The solid (red) lines denote the first-order phase transitions, 
while the dashed (blue) lines and the solid (blue) dot represent  continuous 
phase transitions.}
\label{u0vne0ph}
\end{figure}

With the nearest neighbor interaction $V$ turning on, but the on-site 
repulsion turning off, two solid phases: (0, 2, 2) and (2, 0, 0), emerge, as 
shown in Fig.\ref{u0vne0ph}.
Between the two solids, there are two PSS phases (A and B)
which are characterized by local density fluctuations $\Delta \rho \neq 0$
and the mean PSF order parameter $\phi_p \ne 0$ and the ASF order parameter 
$\phi_a=0$. The boson density in PSS-A $\rho<1$, while that in PSS-B $\rho>1$. 
The PSF phase exists at the outside of the two solids and the 
PSS phases, between the empty phase and the $\rho=2$ MI phase.
Due to the particle-hole symmetry, the phase diagram is symmetric about 
$\mu/V=6$.

The phase diagram is much like that of the meanfield hardcore Bose-Hubbard 
model on the triangular lattice \cite{Murthy}.
The only difference is that the pair states replace the corresponding 
single-atom states: 
the PSS state replaces the ASS state and the PSF state replaces the ASF state.
This can be understood in the following way.
The matrix of the meanfield Hamiltonian of the hardcore bosons on the 
triangular lattice is 
\bea
H_i^{(\rm hc)}=\left(
\begin{array}{cc}
d(0)                        & - t \bar{\phi}_{a,i} \\
-t \bar{\phi}_{a,i} &d(1)
\end{array}
\right),
\label{hcmatrix}
\eea
where the diagonal element is  
\be
d(k)=V (k-\frac{\la n_i \ra}{2}) \bar{n}_i - \mu k+ t \bar{\phi}_{a, i}\phi_{a, i},
\ee
with $k=0, 1$.
Considering $U=0$, the ground state of $H_i$ in Eq. (\ref{t0matrix}) is 
always in the subspace extended by the two Fork states $|0\ra$ and $|2 \ra$.
The matrix of $H_i$ is thus equivalent to $H_i^{(\rm hc)}$ 
after applying the mapping
$2 t_p \to  t, 2 \mu \to \mu $. 
The ground state of $H_i$: 
$|\psi \ra_i=c_{i,0} |0\ra + c_{i,2} |2\ra $ maps to the ground state of 
$H_i^{(\rm hc)}$: $|\psi^{(\rm hc)} \ra_i=c_{i,0} |0\ra + c_{i,2} |1\ra $, 
which leads 
to $\phi_p=\sqrt{2} \phi_a$, with $\phi_p$ the PSF order parameter of $H_i$
and $\phi_a$ the ASF order parameter of hardcore bosons.
Therefore, the phase diagram Fig.~\ref{u0vne0ph} can be obtained from 
the meanfield phase diagram of the hardcore bosons on the triangular 
lattice \cite{Murthy} after applying the mapping.

The properties of phase transitions are also obtained.
Transitions from the PSF to the two solids((0,0,2) and (0,2,2)), and to the 
PSS states are first order, but those from the PSF to the MI phases and 
from the PSS to the two solids are continuous, except for the 
$\mu/V=6$ point marked by a blue dot in Fig.\ref{u0vne0ph}
\cite{Bonnes, Yamamoto}. 
The transition from PSS-A to PSS-B is first-order. 
These results are consistent with what found for hardcore bosons 
on the triangular lattice \cite{tri2,Bonnes, Yamamoto}. 
In a recent work\cite{zhangwz}, a Quantum Monte Carlo study also predicts
the PSS phase between the two solids. However, the PSS region is much smaller
due to quantum fluctuations. 

\subsubsection{$t=0$, $U \neq 0, V \neq 0$}
\label{sec:une0vne0}
\begin{figure}
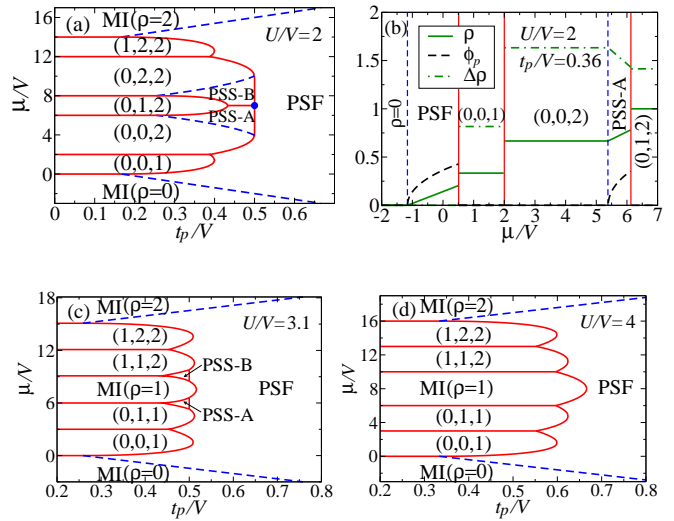

\vskip 0.3cm
\includegraphics[width=4.3cm]{tp-mu-u2-v1.eps}~~
\includegraphics[width=4.1cm]{mu-tp0.36-u2-v1.eps}
\vskip 0.6cm
\includegraphics[width=4.2cm]{tp-mu-u3.1-v1.eps}
\includegraphics[width=4.2cm]{tp-mu-u4-v1.eps}
\caption{(Color online)
(a),(c),(d) Zero-temperature phase diagram for $U/V=2, 3.1, 4$, respectively.
First-order phase transitions are denoted by solid (red) lines,
whereas the continuous phase transitions are represented by dashed (blue) 
lines and the solid (blue) dot.
(b) The density $\rho$, PSF order parameter $\phi_p$, and
solid order parameter $\Delta\rho$ versus $\mu/V$ at $t_p/V=0.36$, $U/V=2$.}
\label{une0vne0ph}
\end{figure}

With the on-site repulsion $U$ present, more phases 
emerge, as shown in Fig. \ref{une0vne0ph}.
When $U/V$ is not large ($<3.3$), the PSS phase persists.
In the strength $U/V>3$, 
a $\rho=1$ MI phase (1, 1, 1) emerges and the solid phases 
(1, 1, 2) and (0, 1, 1) take 
over the (0, 2, 2) and (0, 0, 2) solids. Nevertheless, 
as long as $U/V <3.3$, the 
pair hopping term can still lower the ground state energy by forming the 
PSS-A or B states on the basis of (2, 0, 0) or (0, 2, 2) solid, respectively. 
For example, at $U/V=3.1, \mu/V=6, t_p/V=0.48$, we find the ground state is
a PSS-A state
\bea
|\psi\ra &=& (0.353|0\ra+ 0.935 |2\ra)^{N_A}
(0.912 |0\ra+0.409|2\ra)^{N_B}  \nonumber \\
&&(0.912 |0\ra +0.409 |2\ra)^{N_C},
\eea
which leads to the occupancy (1.750, 0.335, 0.335), ASF order parameter 
$\phi_a=0$, and the mean PSF order parameter $\phi_p=0.508$.
Above the threshold $U/V=3.3$, the PSS phases disappear finally.
The properties of associated phase transitions is  investigated by
checking various order parameters, as illustrated in
Fig. \ref{une0vne0ph}(b),  in which various order parameters are plotted 
as functions of $\mu/V$ at $U/V=2, t_p/V=0.36$.  

\subsection{ Finite $t/t_p$ in the interacting cases}
\label{sec:tvstp}
We now turn to discuss if the above results are stable against putting in
the single-atom hopping $t$.
 
\subsubsection{$V=0,~U \neq 0$}

We first study the change of the phase diagram shown in Fig. \ref{une0v0ph}.

\begin{figure}[htbp]
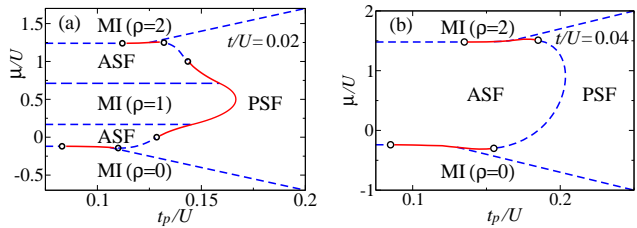

\vskip 0.5cm
\includegraphics[width=4.1cm]{t0.02-tp-mu-u1-v0.eps}~~
\includegraphics[width=3.9cm]{t0.04-tp-mu-u1-v0.eps}
\caption{(Color online)
Zero-temperature phase diagram for $V=0$ at $t/U=0.02$ (a) and
at $t/U=0.04$ (b).
The solid (red) line denotes first-order phase transition,
while the dashed (blue) lines represent continuous phase transitions.
Open circles denote tricritical points. 
}
\label{t02u1v0}
\end{figure}

For small enough $t$, two ASF phases emerge at the boundary of the $\rho=1$ MI 
and the $\rho=2$ MI phases, and at that of the $\rho=1$ MI and the $\rho=0$ 
MI phases, as shown in Fig. \ref{t02u1v0}(a). 
The MI ($\rho=2$)-PSF and the MI ($\rho=0$)-PSF transitions are continuous,
while the MI ($\rho=1$)-PSF transition is still first order.
The MI-ASF transitions and 
the ASF-PSF transition changes from continuous to first order, due to 
the mechanism that levelcorssing preempts the continuous evolving, 
according to our meanfield analysis.
Here the aforementioned coupling of the Ising order parameter and the 
Goldstone mode may change phase transition behavior. Further investigation
beyond the mean-field theory is needed.

When $t$ is large enough, the $\rho=1$ MI phase is excluded by the ASF phase 
completely, as shown in Fig. \ref{t02u1v0} (b). 
Quantum hopping dominates the physics. The phase diagram shows the same
topology as that for non-interacting case (Fig. \ref{ph0}(a)).

\subsubsection{$U=0$ but $V \neq 0$}

\begin{figure}[tpbh]
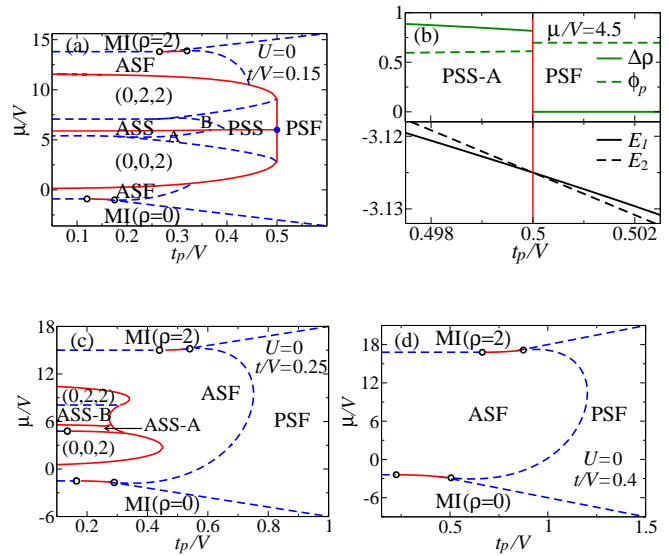

\vskip 0.5cm
\includegraphics[width=4.2cm]{t0.15-tp-mu-u0-v1.eps}~~~
\includegraphics[width=4.0cm]{t0.15-tp-cp4.5-U0-V1.eps}
\vskip 0.7cm
\includegraphics[width=4.2cm]{t0.25-tp-mu-u0-v1.eps}
\includegraphics[width=4.2cm]{t0.4-tp-mu-u0-v1.eps}
\caption{
Zero-temperature phase diagram at $U=0$, $t/V=0.15$ (a), $t/V=0.25$ (c) 
and $t/V=0.4$ (d).
The solid (red) lines denote the first-order phase transitions,
while the dashed (blue) lines and the solid (blue) point represent continuous 
phase transitions.
The  open circles denotes tricritical points.
Levelcrossing and the jumps of order parameters of the PSS-A to PSF transition 
at $\mu/V=4.5$ is shown in (b).
}
\label{u0vne0pht}
\end{figure}

We now check the stability of the phase diagram Fig.~\ref{u0vne0ph} in the 
presence of the single-atom hopping $t$.

With small $t$ present, an ASF region emerges 
between the $\rho=2$ MI and the (0, 2, 2) solid phases, meanwhile another ASF 
phase emerges between the $\rho=0$ MI and the (2, 0, 0) solid phases,  
as shown in Fig. \ref{u0vne0pht}(a). 
In addition, two ASS phases (A: $\rho<1$, B: $\rho>1$)  emerge in the PSS 
region at small $t_p/V$, between two solid phases. Large PSS region persists 
in the region with larger $t_p/V$.
The ASS-PSS transition is continuous and the PSS-PSF 
transition is first order, as illustrated in Fig. \ref{u0vne0pht}(b).

When $t$ becomes stronger, the ASF region is enlarged and encloses the solid 
and the ASS phases, and the PSS phases disappear,
as shown in Fig. \ref{u0vne0pht}(c). 
One interesting phenomenon is that the ASS-A to the (0, 0, 2) solid phase
changes from first order to continuous when $t_p/V$ is weaken,  while the 
ASS-B to the (0, 2, 2) solid phase is always continuous. 

When $t$ becomes even stronger, all solid phases and ASS phases disappear. 
The dominate physics is the ASF-PSF phase transition. The phase diagram 
for $t/V=0.4$ is  presented in Fig. \ref{u0vne0pht}(d), showing the same 
topology as the non-interacting case (Fig. \ref{ph0}(a)). 

\subsubsection{Both $U \neq 0$ and $V \neq 0$}

\begin{figure}[tpbh]
\vskip 0.5cm
\includegraphics[width=8.0cm]{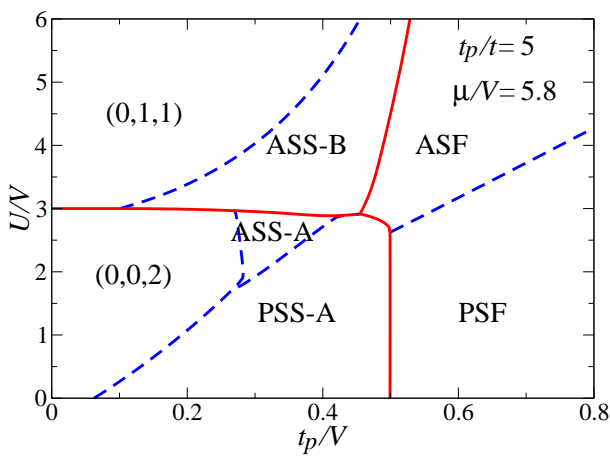}
\caption{(Color online)
Zero-temperature phase diagram at $t_p/t=5, \mu/V=5.8$.
The solid (red) lines denote the first-order phase transitions,
while the dashed (blue) lines represent  continuous phase transitions.
}
\label{tpt5}
\end{figure}

In a real experimental system, the quantum hoppings and atom interactions
can exist simutaneously. The interplay among them results in complex, but
interesting physics effects.
We fix the ratios $\mu/V=5.8$ and $t_p/t=5$ and show the zero-temperature 
phase diagram in the parameter space $t_p/V$ and $U/V$, see Fig. \ref{tpt5}(a).

When the nearest-neighbor repulsion $V$ dominants, the system is in the 
(0, 0, 2) solid phase. As hopping strength grows, the system goes to the 
PSS-A state through a continuous phase transition. 
Further enlarging the hopping strength results in the PSF state through
a first order transition. 
The PSS-A phase is stabilized  in a large parameter region. 

On the other hand, the system is in the (0, 1, 1) solid state, if the on-site 
repulsion $U$ dominants. As hopping strength increases, the system undergoes 
a continuous transition to the ASS-B phase. By increasing further the hopping 
strength, the solid order is destroyed and bosons undergo a first order 
transition to the ASF phase. 
Even larger hopping with the fixed ratio $t_p/t=5$ drives the system into 
the PSF phase through a continuous transition, as expected.

In the middle value of $U/V$ and $t_p/V$, a small region of ASS-A phase exists.

\section{Conclusion}
\label{sec:conc}
We have studied systematically 
the extended Bose-Hubbard model on the triangular lattice, in which the
pair hopping terms are explicitly included and the three-body constraint is 
applied, 
by means of mean-field approaches based on the Gutzwiller ansatz.
The zero-temperature phase diagram and various quantum phase transitions
are investigated in great detail.
In particular, the existence and stability of the pair supersolid phases are 
shown. 
At the limit that the single-atom hopping is zero, we provided 
the mapping between the present model and the hardcore bosons on the 
triangular lattice. The existence of the PSS phase is thus understood.
We have also shown that the PSS phase are stable under the perturbation of 
the on-site repulsion and the finite single-atom hopping.
Experimentally, the triangular optical lattice can be implemented \cite{Becker}.
The three-body constraint can be realized with spin-1 atoms\cite{sp1mott} 
or large three-body loss processes\cite{3bodyloss, 3body2}.
Our results for the pure pair hopping limit ($t=0$) are applicable to 
the system in which bosons pair hopping based on collisions that 
induce transport\cite{cor1}. 
The results in the parameter space $V=0$ are applicable to the atom-molecule 
coupling system proposed  in \cite{xfzhou}, when the three-body constaint is
applied.  
For $V \neq 0$, our results are useful to analyse 
dipolar bosons polarized by an external electric field
and confined in an optical lattice \cite{Sowinski}. However, in such a
system, the occupation-dependent single-atom hopping needs to be discussed.

\acknowledgments
This work is supported by the NSFC under Grant No. 11175018 and No. 11247251.

\end{document}